\documentclass[showpacs]{revtex4}

\usepackage{graphics}
\usepackage{amssymb}
\usepackage{amsmath}

\usepackage{url}
\usepackage[breaklinks=true]{hyperref}
\usepackage{url}
\usepackage[utf8x]{inputenc}

\RequirePackage{color}

\usepackage{amsfonts}
\usepackage{graphicx,graphics}
\usepackage{latexsym}
\usepackage{epstopdf,subfigure}
\begin{document}
%
\title{On Dark Matter As A Geometric Effect in the Galactic Halo}


\author{Ugur Camci} \email{ucamci@rwu.edu,ugurcamci@gmail.com}

\bigskip

\affiliation{Department of Chemistry and Physics, Roger Williams University, One Old Ferry Road, Bristol, RI 02809, USA\label{addr1} }

\bigskip

\date{\today}

\begin{abstract}
We obtain more straightforwardly some features of dark matter distribution in the halos of galaxies by considering the spherically symmetric space-time, which satisfies the flat rotational curve condition, and the geometric equation of state resulting from the modified gravity theory. In order to measure the equation of state for dark matter in the galactic halo, we provide a general formalism taking into account the modified $f(X)$ gravity theories. Here,  $f(X)$ is a general function of $X \in \{ R, \mathcal{G}, T \}$, where $R, \mathcal{G}$ and $T$ are the Ricci scalar, the Gauss-Bonnet scalar and the torsion scalar, respectively. These theories yield that the flat rotation curves appear as a consequence of the additional geometric structure accommodated by those of modified gravity theories. Constructing a geometric equation of state $w_{_X} \equiv p_{_X} / \rho_{_X}$ and inspiring by some values of the equation of state for the ordinary matter, we infer some properties of dark matter in galactic halos of galaxies.
\end{abstract}



\maketitle

\section{Introduction}

While the evolution of the universe is driven by the dark energy (DE), the formation of galaxies should be mainly determined by the dark matter (DM) which provides the potential wells where baryons collapse to originate the visible component of galaxies. The direct evidence of DM and DE follows from distance measurements of type Ia supernovae (SNe Ia) indicating that the expansion of the universe is speeding up \citep{riess99,perl98}, which is the most important evidence of a breakdown of general relativity (GR). The accelerated expansion of the universe can be explained by different models of gravity than GR, where the Einstein-Hilbert Lagrangian is generalized by some general functions such that $f(R), f(T), f(R,\mathcal{G}),$ etc. The cosmic microwave background (CMB) anisotropy measurements are equally strong indirect evidence for that type of matter and energy. All of the projects to probe DE have the common feature of surveying wide areas to collect large samples of galaxies, clusters or supernovae. An additional indirect evidence for DM comes from the connection between galaxies/clusters and their halos, and for that type of structures, much higher mass estimates than expected from the observed stars and gas indicate a breakdown of Newton's theory of gravity \cite{wt2018}. The modified gravity models, which are commonly considering as gravitational alternative for DM, can explain the extra matter content to be needed in addition to the observational data coming from galaxies. So the gravitational properties of galaxies and clusters have to be checked by considering the modified theories of gravity.

The equation of state (EoS) for the ordinary matter, $ w = p / \rho $, can take an arbitrary value between $-1$ and $1$, where $w = -1$ corresponds to a cosmological constant, which is also matched to an exterior Schwarzschild vacuum solution, $w = 0$ for dust matter and $w=1$ for a stiff matter. The EoS parameter $w=1/3$ corresponds to the relativistic matter (the radiation), like photons and massless neutrinos. For quintessence models the EoS parameter range is $-1< w < -1/3$, and the matter with the property $w < -1$ is dubbed the phantom energy. Due to the apparent abundances of DM in  most of the galactic halos, we can suggest a geometric EoS of the form  $w_X \equiv p_X / \rho_X$ to the spherically symmetric space-times for studying gravitational properties of DM in the galactic halo. Then, using the field equations of modified gravity model and ignoring the contribution of ordinary baryonic matter to that field equations, one can write those of field equations in terms of $\rho_X$ and $p_X$ which are $X \in \{ R, T, \mathcal{G},... \}$ term contributions to energy density and pressure, where $R, T$ and $\mathcal{G}$ are the Ricci scalar, the torsion scalar and the Gauss-Bonnet scalar which is defined by $\mathcal{G} = R^2 - 4 R_{\mu \nu} R^{\mu \nu} + R_{\alpha \beta \mu \nu} R^{\alpha \beta \mu \nu} $ with $R_{\mu \nu}$ and $R_{\alpha \beta \mu \nu}$ being the Ricci tensor and the Riemann tensor, respectively. There are some studies of trying to measure the EoS of DM in the galactic halo where it has been considered some halo models of galaxies predicting a significant amount of pressure in the DM fluid such that some variations of scalar field DM \cite{bk2003,mgu2000,mgn2000,p2000,als2003}, or string fluid \cite{s1995}.

The theoretical expectations for large amounts of luminous matter in galaxies are not in agreement with observations. An insufficient amount of standard matter in galaxies is the main problem, and the existence of some non-luminous matters, the DM, should be necessary to explain both dynamics and structure formation of galaxies. It is well known that the extended flat rotation curves of spiral galaxies provide some of the strongest evidence for mass discrepancies in galaxies, and are usually interpreted as evidence for DM halos. The observed flat rotational curves in outskirts of galaxies are usually interpreted as evidence for the DM halos, i.e. most of the DM lies in the halos of galaxies in which the rotation curves are strictly flat \cite{sellwood1999}. According to the current information for galaxy formation, every galaxy forms and evolves within a DM halo, and it is expected that visible galaxies are hosted by all halos. The aim of this paper is to point out the main astrophysical consequences that arise by considering the modified theories of gravity in terms of a different perspective of the DM included in galactic halo. Specially, inspiring by some values for the EoS parameter which is used to the ordinary matter, we will investigate the kind of DM in halos of galaxies.

The outline of this paper is as follows. In the following section, we briefly describe the motion of massive test particles under the conditions for circular orbits. In Section \ref{sec:mgt}, we present the form of the modified gravity theory $f(X)$ , where $X \in \{ R, \mathcal{G}, T,... \}$, as well as write down its equations of motion for each of $f(R)$, $f(R,G)$ and $f(T)$ gravity theories that yields constant rotational curves in the outskirts of galaxies. Finally, we conclude with summary and discussion of the obtained results in Section \ref{sec:discuss}.

\section{The Motion of Test Particles}
\label{sec:motion}

In order to obtain results which are relevant to the galactic dynamics, we assume that the galactic halo has spherical symmetry and that dragging effects on material particles (stars and dust) are inappreciable. Therefore, we restrict our study to the static and spherically symmetric metric. The most general static and spherically symmetric metric can be written as
\begin{equation}
{\rm d}s^2 = - A (r) {\rm d}t^2 + B (r) {\rm d}r^2 + d\Omega^2 \, , \label{metric}
\end{equation}
where $d\Omega^2 = C(r) \left( {\rm d}\theta^2 + \sin^2\theta {\rm d}\varphi^2 \right)$ and ($r$,$\theta$,$\varphi$) are the spherical coordinates. Then, the equations of motion for a test particle in the space-time \eqref{metric} can be derived from the Lagrangian
\begin{equation}
2 \mathcal{L} = - A(r) \dot{t}^2 + B(r) \dot{r}^2 + C(r) \left[ \dot{\theta}^2 + \sin \theta^2 \dot{\varphi}^2 \right], \label{lagr}
\end{equation}
where a dot means derivative with respect to the proper time. From the above Lagrangian \eqref{lagr}, the generalized momenta become
\begin{eqnarray}
& & p_t = - E = - A \dot{t} \, , \,\, p_r = B \, \dot{r} \, , \nonumber \\ & &   p_{\theta} = L_{\theta} = C \, \dot{\theta} \, ,  \,\,  p_{\varphi} = L_{\varphi} =  C \, \sin^2 \theta \, \dot{\varphi}, \label{gmom}
\end{eqnarray}
where $E$ is the total energy of a test particle and $L_i$ are the components of its angular momentum. Using $L^2 = L_{\theta}^2 + L_{\varphi}^2 / \sin^2 \theta$ which is the first integral corresponding to the squared total angular momentum, the norm of the four-velocity ($g_{\mu \nu} u^{\mu} u^{\nu} = -1$) yields
\begin{equation}
E^2 = V(r) + A B \, \dot{r}^2 \, , \label{E2}
\end{equation}
where $V(r) \equiv A ( 1 + L^2 / C)$ is the effective potential. Thus, the conditions for circular orbits $\dot{r} = 0$ and $\partial_r V(r) = 0$ lead to
\begin{equation}
E^2 = \frac{ A^2 \, C'}{ A \, C' - C \, A'}, \qquad L^2 = \frac{ C^2 \, A'}{ A \, C' - C \, A'} \, . \label{E2-L2}
\end{equation}
The definition of tangential velocity of a test particle is given by \cite{bohmer2008}
\begin{equation}
v_{tg}^2 = \frac{C}{A} \left( \frac{ d \Omega}{ d t} \right)^2  = \frac{C}{A} \frac{\dot{\Omega}^2}{\dot{t}^2} = \frac{A}{C} \frac{L^2}{E^2}   \, . \label{def-vtg2}
\end{equation}
Then, using the constants of motion \eqref{E2-L2}, it follows that the tangential velocity of this test particle is
\begin{equation}
v_{tg}^2 = \frac{ A'/A}{C'/C} \, , \label{vtg2}
\end{equation}
which has the form
\begin{equation}
v_{tg}^2 = \frac{r A'}{2 A} \, , \label{vtg2-2}	
\end{equation}
for $C(r) = r^2$, where a prime represents derivative with respect to $r$. Since the motion of test particles is defined via the geodesic equations, the relation \eqref{vtg2} is independent of any theory of gravity. In the flat rotation curves region, where $v_{tg} \approx {\rm constant}$, integration of Eq.\eqref{vtg2} gives
\begin{equation}
A(r) = A_0 C(r)^{v_{tg}^2} \, , \label{A-vtg2}
\end{equation}
which takes the form
\begin{equation}
A(r) = A_0 r^{\ell}  \, , 	
\end{equation}
for $C(r) = r^2$, where $A_0$ is an integration constant and $\ell = 2 v_{tg}^2$. Thus, the  metric \eqref{metric} with the relation \eqref{A-vtg2} describes the geometry of the space-time for the DM dominated regions where the test particles move in constant rotational curves \cite{mgn2000,bohmer2008}.

\section{Modified Gravity Models for Flat Rotation Curves Region }
\label{sec:mgt}
The generic action that we will consider for modified gravity theories reads
\begin{equation}
\mathcal{S} = \int{ d^4 x \sqrt{-g} \left[ \frac{1}{2 \kappa^2} f(X) + \mathcal{L}_{\rm m} \right] } \, , \label{action}
\end{equation}
where $\kappa^2 = 8 \pi G_N$ is the standard gravitational coupling, $g= \det ( g_{\mu \nu} )$, $\mathcal{L}_{\rm m}$ is the ordinary matter Lagrangian and the function $f$ depends on $X \in \{ R, \mathcal{G}, T,... \}$. We also point out that the galaxy is embedded in a modified spherically symmetric geometry, generated by the non-zero contributions of the modified gravitational action. Hereafter, in order to obtain effects of the DM in the halos of galaxies, we have assumed that there exists no the luminous matter, i.e. $\mathcal{L}_{\rm m}=0$. Before obtaining the field equations by varying the Lagrangian according to the coefficients of metric \eqref{metric}, we note the fact that the metric variable $B$ does not contributes to dynamics by using a point-like Lagrangian approach which implies that the Lagrangian does not depends on the generalized velocity $B'$, but the equation of motion due to the variation of $B$ has to be considered as a further constraint equation \cite{capo2007}. Under the latter fact, the action for the geometrical part has the form
\begin{equation}
\mathcal{S}_{f(X)} = \int \mathcal{L}_{f(X)} (A,B,C,X,A',C',X') dr \, .
\end{equation}
Then, taking variation of this action with respect to the metric coefficients $A, B$ and $C$ of the static spherically symmetric space-time \eqref{metric}, one can write the field equations as
\begin{eqnarray}
& &  F_ 1 (q^i, q'^i, A'', C'') = \rho_{_X} \, , \label{feq1} \\ & &  F_2 (q^i,q'^i, A'', C'') = p_{_X} \, , \label{feq2} \\ & & p'_{_X} + F_3 ( q^i,q'^i) ( \rho_{_X} + p_{_X} ) = Q (q^i,q'^i,X,X') \, , \label{feq3}
\end{eqnarray}
where $F_i $ are the functions of the metric coefficients $q^i = \{A, B, C \}$ and their first and/or second derivatives, $\rho_{_X} (q^i,q'^i,X,X',X'')$ and $p_{_X} (q^i,q'^i,X,X',X'')$ are the density and pressure for the modified gravity theory, respectively. Then, inspiring by some values for the EoS parameter used in the ordinary matter, the geometric EoS defined by $w_X \equiv p_X / \rho_X$ has capacity to inform us about the kind of DM in the galactic halos. For instance, if the modified gravity model gives rise to the geometrical EoS parameter as $w_{X} = 1/3$, then it may be corresponding to the {\it dark radiation} \cite{ackerman2009} inside the galactic halo. Now, we will take three examples to explain the above theoretical construction of the extended gravity model that yields constant rotational curves in the outskirts of galaxies even though it can be given so much examples of the idea represented here.

\begin{itemize}
	
\item {$f(R)$ gravity:}
Considering the background metric \eqref{metric}, the point-like Lagrangian for $f(R)$ gravity is given by \cite{capo2007}
\begin{eqnarray}
& & \mathcal{L}_{f(R)} =  f_R \left( \frac{ A' C'}{\sqrt{A B}} +  \sqrt{\frac{A}{B}}\frac{C'^2}{2 C} \right)  + f_{RR} R' \left( \frac{C A' }{\sqrt{A B}}  + 2\sqrt{\frac{A}{B}}  C'  \right)  + \sqrt{A B} \left[ C f  + (2- C R) f_R \right]  .  \label{lagr1}
\end{eqnarray}
After varying \eqref{lagr1} with respect to $A, B$ and $C$, the field equations have the form of \eqref{feq1}-\eqref{feq3}, which read
\begin{eqnarray}
& & \frac{1}{B} \left( \frac{B' C'}{B C} - \frac{2 C''}{C} + \frac{C'^2}{2 C^2} + \frac{2 B}{C} \right) = \rho_{_R}, \\ & & \frac{1}{B} \left( \frac{B' C'}{B C}  + \frac{C'^2}{2 C^2} - \frac{2 B}{C} \right) = p_{_R}, \\& & f_R \Big[ \frac{A''}{A} + \frac{C''}{C} - \frac{A' B'}{2 A B} + \frac{C'}{2 C} \left( \frac{A'}{A} - \frac{B'}{B} \right) - \frac{A'^2}{2 A^2} - \frac{C'^2}{2 C^2}  \Big]  + f'_R \left( \frac{A'}{A} - \frac{B'}{B} + \frac{C'}{C} \right) \nonumber \\ & & \qquad + 2 f''_R + B ( R f_R - f) = 0,
\end{eqnarray}
where $\rho_{_R}$ and $p_{_R}$ are defined by
\begin{eqnarray}
& & \rho_{_R} \equiv \frac{1}{f_R} \Big( R f_R - f  + \frac{1}{B} \left[ f'_{R} \left( \frac{2 C'}{C} - \frac{B'}{B} \right) + 2 f''_{R} \right] \Big) \, ,  \\ & &  p_{_R} \equiv \frac{1}{f_R} \left[ f- R f_R - \frac{ f'_R}{B} \left( \frac{A'}{A} + \frac{2 C'}{C} \right) \right] \, ,
\end{eqnarray}
where $f'_R = f_{RR} R'$ and $f''_R = f_{RR} R'' + f_{RRR}R'^2$. For an exact solution of the power law gravity $f(R) = f_0 R^n$, the metric coefficients are given by
\begin{eqnarray}
& & A(r) = A_0 r^{\ell}\, , \qquad  B(r) = \frac{(1 + 2 n -2 n^2 )( 4 n^2 - 10 n +7)}{(n -2)^2} \, ,   \qquad C(r) = r^2   \, , \label{sol1-A-B}
\end{eqnarray}
with $\ell = 2 (n-1)(2n-1)/(2 - n)$.  Therefore, the tangential velocity for this case becomes
\begin{equation}
v_{tg}^2 = \frac{(n-1)(2n-1)}{(2 - n)},  \label{vtg-1}
\end{equation}
due to the relation $\ell = 2 v_{tg}^2$. Here, the density and pressure of $R^n$ gravity have the form
\begin{equation}
\rho_{_R} (r) = \frac{\rho_0}{r^2} \, , \qquad   p_{_R} (r) =  \frac{p_0}{r^2} \, ,
\end{equation}
where $\rho_0$ and $p_0$ are found as
\begin{eqnarray}
& & \rho_0 = \frac{ 2 (n-1) ( 8 n^3 - 20 n^2 + 11 n + 3 ) }{ (2 n^2 - 2 n -1)(4 n^2 - 10 n + 7)} \, , \\  & &   p_0 = \frac{ 2 (1-n)( 8 n^3- 24 n^2 + 21 n -1 ) }{ (2 n^2 - 2 n -1)( 4 n^2 -10 n + 7)} .
\end{eqnarray}

The relation \eqref{vtg-1} and the form of $\rho_{_R}$ and $p_{_R}$ explicitly represent that the constant tangential velocity $v_{tg}$, the density $\rho_{_R}$ and pressure $p_{_R}$ vanish as $n=1$, which is the GR case, that is, it can be concluded that the GR does not give any information about the DM dominated region of galaxies. Furthermore, the geometric EoS $w_{_R} = p_{_R} / \rho_{_R}$ is constant, and found to be
\begin{equation}
w_{_R} = -\frac{( 8 n^3- 24 n^2 + 21 n -1 )}{(8 n^3 - 20 n^2 + 11 n + 3)} \, . \label{w-sol1}
\end{equation}

\item $f(R,\mathcal{G})$ gravity: This theory of gravity is a generalization of $f(R)$ gravity containing a Gauss$-$Bonnet scalar $\mathcal{G}$ in the action \eqref{action}. For this case, the point-like Lagrangian for the static spherical symmetric metric \eqref{metric} takes the form \cite{bdc2020}
\begin{eqnarray}
& & \mathcal{L}_{f(R,\mathcal{G})} = \mathcal{L}_{f(R)} +   \frac{f_{\mathcal{G}\mathcal{G}} \mathcal{G}'}{ \sqrt{A B}}  \left( 4 A' - \frac{A' C'^2}{ B C} \right)  - \mathcal{G} \sqrt{A B} f_{\mathcal{G}} \,  ,  \label{lagr2}
\end{eqnarray}
where $\mathcal{L}_{f(R)}$ is as given by \eqref{lagr1}. Then, solving the field equations obtained from the above Lagrangian for $f(R, \mathcal{G}) = f_0 R + f_1 \sqrt{\mathcal{G}}$ gravity, it is found the following exact spherically symmetric solution \cite{bdc2020} :
\begin{equation}
A(r) = A_0 r^{\ell}, \qquad B(r) = \frac{4 - \ell^2}{4}, \quad C(r) = r^2  \, , \label{sol2}
\end{equation}
where $f_1 = - 2 f_0 \ell \sqrt{ 2 \ell(\ell-2)} / ( \ell^2 - 2 \ell + 8)$,  and the density and pressure are
\begin{equation}
\rho_{_{R\mathcal{G}}} (r) = \frac{ 2 \ell^2}{( \ell^2 -4) \, r^2},  \quad  p_{_{R\mathcal{G}}} (r) = -\frac{ 2 \ell (\ell + 4)}{( \ell^2 -4)\, r^2}  \, , \label{sol2-rho-p}
\end{equation}
with the geometric EoS $w_{_{R\mathcal{G}}}  = - 1 - 4 / \ell$ which reduces to $w_{_{R\mathcal{G}}}  = - 1 - 2 v_{tg}^{-2}$ since the relation $\ell = 2 v_{tg}^2$.

\bigskip

\item $f(T)$ gravity: \\
In this gravity theory, the gravitational contributions to the
metric tensor become a source of torsion $T$ rather than curvature $R$, i.e. the connection considered in this theory of gravity is distinct from the regular Levi-Civita connection which is replaced with its Weitzenb\"{o}ck analog. In order to be consistent with the study \cite{bu2019}, we will use the signature $(+,-,-,-)$ for spherically symmetric metric which has the form $ds^2 = A(r)^2 dt^2 - B(r)^2 dr^2 - C(r)^2 (d\theta^2 + \sin \theta^2 d\varphi^2)$. The Lagrangian $\mathcal{L}_{f(T)}$ for the latter metric is given by  \cite{bu2019}
\begin{eqnarray}
& & \mathcal{L}_{f(T)} = A B  \left[  C^2 ( f - T f_T ) + 2 f_T \right] +  2 f_T A C \Big[ \frac{A C'}{B} \left( \frac{2 A'}{A} +  \frac{C'}{C}  \right)  - 2 \left( \frac{A'}{A} + \frac{C'}{C} \right) \Big]  \, .\qquad \label{lagr3}
\end{eqnarray}
After varying the above Lagrangian with respect to $A, B$ and $C$, the field equations for power-law $f(T)= f_0 T^n$ gravity can be written in the form of \eqref{feq1}-\eqref{feq3} as follows
\begin{eqnarray}
& & \frac{2}{B^2} \left( \frac{ 2 B' C'}{B C} - \frac{2 C''}{C} - \frac{C'^2}{C^2} + \frac{B^2}{C^2}   \right) = \rho_{_T},  \label{fe1} \\ & & \frac{2}{B^2} \left( \frac{ 2 A' C'}{A C} + \frac{C'^2}{C^2} - \frac{B^2}{C^2}   \right) = p_{_T},  \label{fe2} \\ & & p'_{_T} + \left( \frac{A'}{A} + \frac{C'}{C} \right) ( \rho_{_T} + p_{_T} ) + 4 (n-1) \frac{ A' C' T'}{ A B^2 C T} = 0 \, ,  \label{fe3}
\end{eqnarray}
where $\rho_{_T}$ and $p_{_T}$ are defined as
\begin{eqnarray}
& & \rho_{_T} \equiv (n-1) \left[ \frac{T}{n} + \frac{4}{C B^2} \frac{T'}{T} \left( C' - B \right)  \right] \, ,  \qquad  p_{_{T}} \equiv \frac{(1-n)}{n} T  \, , \label{def-rho-p-T}
\end{eqnarray}
which are the torsion contributions to energy density and pressure. If $n=1$, then the theory becomes the Teleparallel Equivalent of General Relativity (TEGR), and the energy density and pressure given by \eqref{def-rho-p-T} vanish, just as expected. Using the field equations for $f(T)= f_0 T^n$ gravity, where $n \neq 0, 1, \frac{1}{2}, \frac{5}{6}, \frac{5}{4}, \frac{3}{2}$, it is reported the following exact solution in Ref. \cite{bu2019}
\begin{eqnarray}
& & A(r) = A_0 r^{\ell} \, , \nonumber \\ & & B(r) = \frac{( 2n-1)(4n -1)}{(4 n^2 - 8 n + 5)} \, ,  \label{sol3-A-B} \\  & &  C(r) = r   \, , \nonumber
\end{eqnarray}
where $\ell = 4n (n-1) (2n-3)/( 4n^2 - 8n + 5)$. Here it is found the relation $\ell = v_{tg}^2$ because of the form of metric coefficients in this case, which yields that the tangential velocity has the form
\begin{equation}
v_{tg}^2 = \frac{4 n (n-1)(2n-3)}{4 n^2 - 8 n + 5}.  \label{vtg-2}
\end{equation}
For this solution, the density and pressure of $T^n$ gravity become
\begin{equation}
\rho_{_T} (r) = \frac{\rho_1}{r^2} \, , \qquad  p_{_T} (r) = \frac{p_1}{r^2} \, ,
\end{equation}
where $\rho_1$ and $p_1$ are obtained as
\begin{eqnarray}
& & \rho_1 = \frac{ 8 n (n-1) (2n-3)(6n-5) }{ (2n-1)^2 (4n -5)^2 }, \qquad  p_1 = \frac{ 8 n (n-1)(2n-3)^2 }{ (4n - 5)(2n -1)^2 } \, ,
\end{eqnarray}
which gives a constant geometric EoS $w_{_T} = p_1 / \rho_1$ as
\begin{equation}
w_{_T} = \frac{(2n -3)(4n-5)}{6n-5}.  \label{w-sol3}
\end{equation}

\end{itemize}

----------------------------
\begin{figure*}[t]
	\centering
	\includegraphics[scale=0.44]{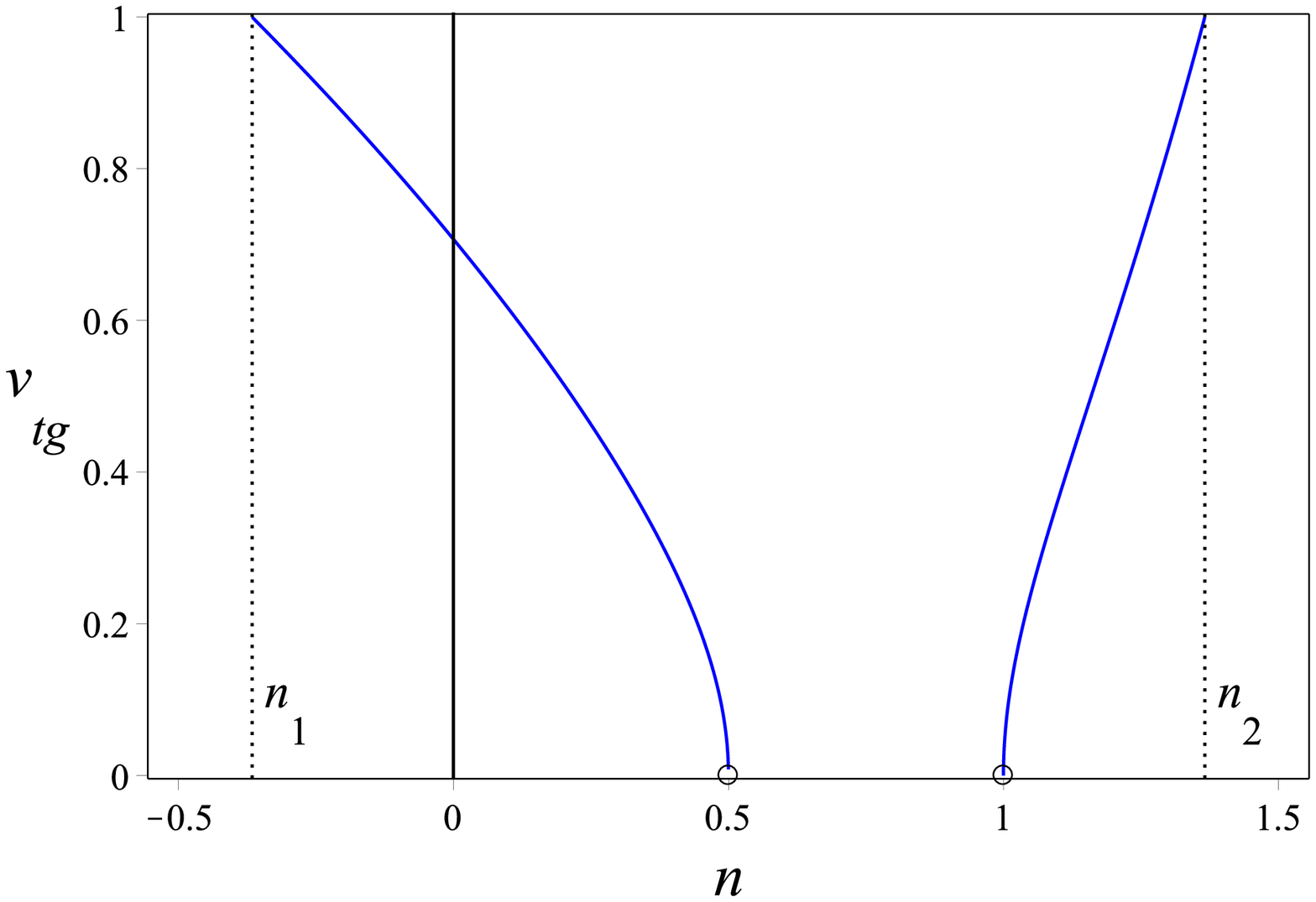}
	\includegraphics[scale=0.44]{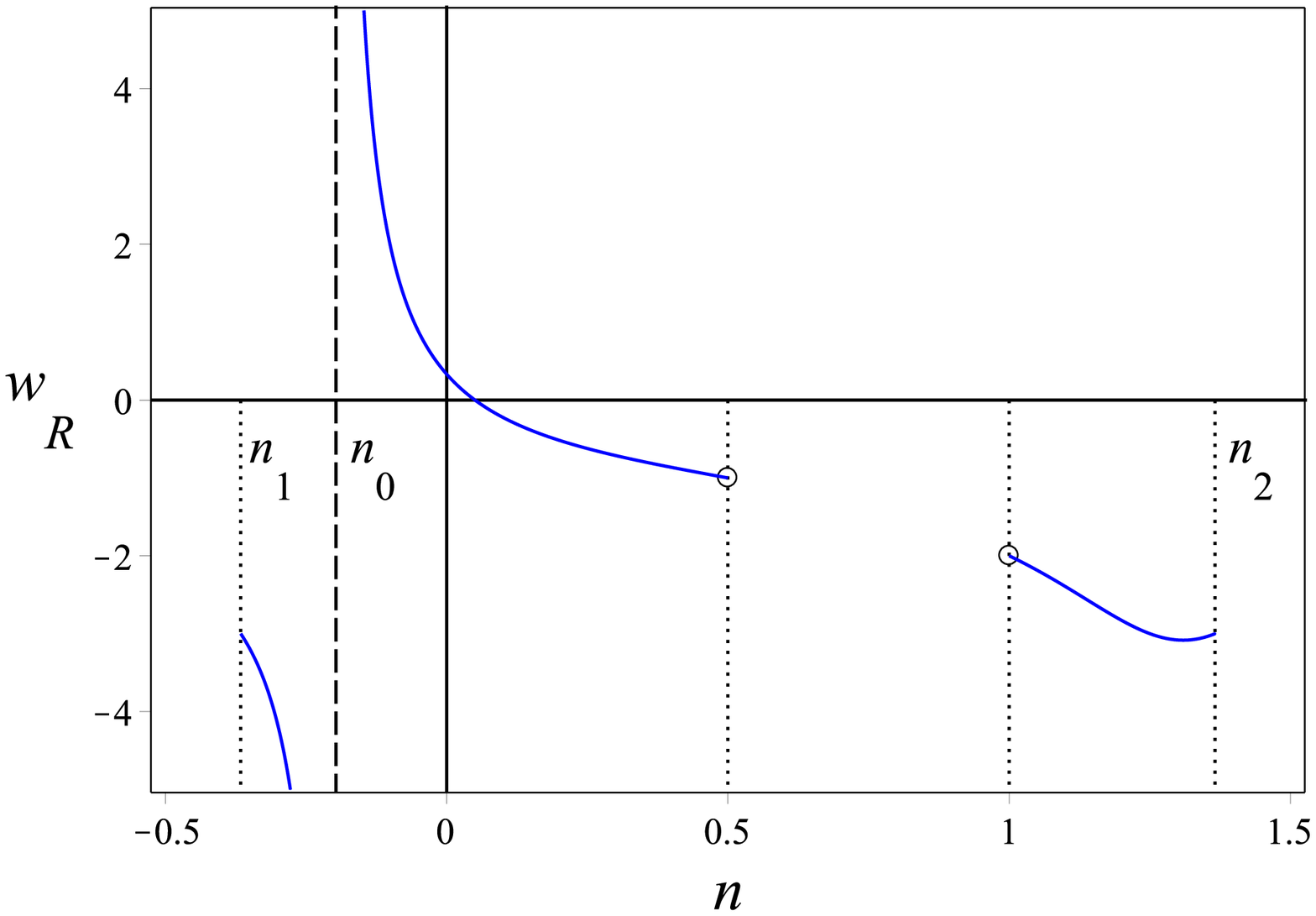}
	\caption{ Plots showing the tangential velocity $v_{tg}$ (left panel) and the geometrical equation of state $w_{_R}$ (right panel) given in \eqref{w-sol1} vs the power $n$ of $R^n$ gravity where $n_0 = -0.1967807237, \, n_1 = (1 -\sqrt{3})/2$ and $n_2 = (1+ \sqrt{3})/2$.}
	\label{fig1}
\end{figure*}
----------------------------

\section{Discussions and conclusions}
\label{sec:discuss}

For the three examples considered in the previous section, the metric coefficients have the form
\begin{equation}
 A(r)= A_0 r^{\ell}, \quad  B(r)=B_0 ={\rm constant} \, , \label{metric-final}
\end{equation}
and $ d\Omega^2 = r^2 \left( {\rm d}\theta^2 + \sin^2\theta {\rm d}\varphi^2 \right)$. In all examples for the considered $f(X)$ theory of gravity, where $X \in \{ R, \mathcal{G}, T\}$, the geometric density $\rho_X$ and pressure $p_X$ with a spheroidal profile, i.e. $C(r) = r^2$, are found as isothermal, which means that they are proportional with distance as $1/r^2$, and the geometrical EoS $w_X$ is constant. In the first and third examples discussed above section, we have shown that there is a connection between power of the modified gravity theory and the corresponding geometric EoS parameter describing the intrinsic property of the DM halos of galaxies. In the second example, we have taken an extension of GR in which an extra term included proportional with the square root of Gauss-Bonnet scalar $\mathcal{G}$, and found that the geometric EoS is directly related with flat rotational velocity of galaxies.

\bigskip

In order to test the consistency of results, it has been considered the Newtonian limit of the model by \cite{bohmer2008}, in which they have restricted their analysis to the constant velocity region, and found that the model has a well-defined Newtonian limit. In the Newtonian limit the $g_{\rm tt}$ component of the metric tensor has the form $A \approx 1 + 2 \Phi_N$, where $\Phi_N$ is the Newtonian gravitational potential, and it is obtained for the constant velocity regions around galaxies in the limit of large $r$ as follows \cite{bohmer2008} :
\begin{equation}
  \Phi_N (r) \approx  v_{\rm tg}^2 \ln \left( \frac{r}{r_0} \right)  \, ,  \label{potential}
\end{equation}
where we take $A_0$ and $\ell$ as $A_0 = r_0^{-\ell}$ and $\ell = 2 v_{\rm tg}^2$ in \eqref{metric-final}. Thus, Eq. \eqref{potential} represents that the correction term to the Newtonian potential in the DM dominated region, where the rotation curves are strictly flat, must have a logarithmic dependence on the radial coordinate $r$.

\bigskip

All test particles in stable circular motion move at the speed of light when $v_{tg} = 1$, but this gives rise to a contradiction by observations at the galactic scale. Furthermore, the tangential velocity $v_{tg}$ tends to zero in the limit of large $r$. Thus, the tangential velocity $v_{tg}$ has to be at the interval $0 < v_{tg} < 1$. During early epochs of the universe, the DM velocity is not so small than the speed of light, for example the relative velocity is $v_{rel} \propto 0.3 c$ at freeze-out epoch \cite{ty2018}.  While at later times such as the DM halos today and during the recombination epoch, the DM velocity is very smaller. For instance, the tangential velocities of DM halos in dwarf galaxies, spiral galaxies and galaxy clusters are approximately proportional with $10^{-5}, 10^{-3}$ and $10^{-2}$ in units of $c=1$ at distances large enough from the galactic center, respectively. In the first and third examples, it is not only found all the spectrum of flat rotation curves but also the geometric EoS for the power law $R^n$ and $T^n$ theories of gravity at the interval $0 < v_{tg} < 1$. Now we conclude our findings in three examples as follows:

\begin{itemize}
\item $f(R)= f_0 R^n$ gravity:
    The relation between $v_{tg}$ and the power $n$, the Eq. \eqref{vtg-1}, is a second order algebraic equation for $n$, and it has a solution as
    \begin{equation}
    n = \frac{1}{4} \left( - v_{tg}^2  + 3 \pm \sqrt{ v_{tg}^4 + 10 v_{tg}^2 +1} \, \right) \, .
    \end{equation}
    So, we can exactly calculate $n$ for some specific tangential velocities. For instance, if $v_{tg} = 10^{-3}$, which is the rotational velocity of spiral galaxies, then we find $n= 1.000001$ or $n=0.4999985$. For $n= 1.000001$,  the geometrical density and pressure coefficients $\rho_0$ and $p_0$ become $\rho_0 = -4 \times 10^{-6}$ and $p_0 = 8 \times 10^{-6}$, while for $n=0.4999985$ we find $\rho_0 = 1.000001$ and $p_0 = -0.999999$. The latter result is most important in  the sense that the galactic halos for spiral galaxies prefer almost {\it dark energy} with a geometric EoS $w_{_R} = -0.999998$, which indicates that $n=0.4999985$ for the flat rotational velocity region. A general picture for the relation between $v_{tg}$ and the power $n$ is shown in Fig. \ref{fig1} which gives us some important information on the DM halos of galaxies. It is seen from this figure that $n \in [ n_1, 0.5 ) \cup ( 1, n_2 ]$. For $n=-0.0578084147$ and $n=0.0504825997$ we arrive the {\it dust DM} ($w_{_R} = 0$) and {\it stiff DM} ($w_{_R} = 1$), respectively. Also, for $n=n_1$ and $n=n_2$ that gives $v_{tg}=1$, the geometric EoS yields $w_{_R} = -3$ which is an EoS parameter for {\it the phantom energy} ($w_{_R} < -1$). At the interval $n_1 < n < 0.1321342251$, one get the quintessence EoS parameter values, which are in the range $-1< w_{_R} < -1/3$. In this model, the dark radiation ($w_{_R} = 1/3$) appears if $n=0$, for which $v_{tg} = 0.7071067812$. For $n= n_0 = -0.1967807237$, the tangential velocity is $v_{tg} = 0.8713186295$ and $w_{_R}$ tends to the infinity, which represents that there is a phase transition at this evolution stage of galaxy.
    Using the observed best-fit values of $v_{\rm flat}$ to the flat rotation curves of sample galaxies \cite{frank2016,cp1998,kamamda2017},  we present the tangential velocity of test particles, the corresponding power $n$, and the equation of state parameter $w_R$ for $R^n$ gravity in Table I.

    Capozziello et al. (2007b) have investigated the possibility that the observed flatness of the rotation curves of spiral galaxies for $R^n$ gravity is not evidence for the existence of DM, in which they have obtained the corrected gravitational potential as
    \begin{equation}
      \Phi(r) = -\frac{G m}{2 \, r} \left[ 1 + \left( \frac{r}{r_c} \right)^{\beta} \right] \, , \label{pot-Capo}
    \end{equation}
    where $m$ is a pointlike mass, $r_c$ is an arbitrary parameter, and $\beta$ is a parameter depending on the power $n$. The above potential is still asymptotically decreasing, but the corrected rotation curve is higher than the Newtonian one and not flat. Because of that the Newtonian potential for the DM dominated region should have a logarithmic dependence such as Eq. \eqref{potential}, the correction term to the Newtonian potential in the DM dominated region, where the rotation curves are flat, does not appear in Eq. \eqref{pot-Capo}. The difference in the Newtonian limit in our model results in different values of the parameter $n$ for $R^n$ gravity.

\begin{table*}
\caption{\label{Tab1} Properties of sample galaxies in $R^n$ gravity. An explanation of the quantities appeared in the columns: $v_{\rm flat}$ - galactic flat rotation velocity data, $v_{\rm tg}$-tangential velocity, $c$-velocity of light, $n$-power of $R^n$, $w_R$- equation of state parameter. }
\begin{tabular}{ccccc}
Galaxy & $v_{\rm flat} ({\rm km \, s^{-1}})$  & $v_{\rm tg} = v_{\rm flat} / c$ ( $ 10^{-3} $ ) & $n$ & $w_R = p_0 / \rho_0$ \\
\hline
Milky Way & $200$ & $0.667$  & $0.4999993327$ & $-0.9999991107$ \\ & & & $1.000000445$ & $-2.000001558$ \\ \\
NGC2841   & $310$ & $1.034$  & $0.4999983963$ & $-0.9999978618$ \\ & & & $1.000001069$ & $-2.000003747$ \\ \\
NGC3521   & $231$ & $0.77$   & $0.4999991107$ & $-0.9999988142$ \\ & & & $1.000000593$ & $-2.000002081$ \\ \\
NGC3198   & $152$ & $0.507$  & $0.4999996144$ & $-0.9999994856$ \\ & & & $1.000000257$ & $-2.000000894$ \\ \\
NGC2403   & $97$  & $0.324$  & $0.4999998424$  & $-0.99999979$ \\ & & & $1.000000105$ & $-2.000000368$ \\ \\
UGC128    & $140$ & $0.47$   & $0.4999996687$ & $-0.9999995582$ \\ & & & $1.000000221$ & $-2.000000768$ \\ \\
UGC5721   & $80$  & $0.267$  & $0.4999998931$ & $-0.9999998578$ \\ & & & $1.000000071$ & $-2.000000244$ \\ \\
UGCA442   & $57$  & $0.19$   & $0.4999999459$ & $-0.9999999282$ \\ & & & $1.000000036$ & $-2.000000121$ \\ \\
DDO154    & $45$  & $0.15$   & $0.4999999663$ & $-0.9999999551$  \\ & & & $1.000000022$ & $-2.000000072$ \\
\hline
\end{tabular}
\end{table*}

\item $f(R,\mathcal{G}) = f_1 R + f_2 \sqrt{\mathcal{G}}$ gravity:

In Ref. \cite{mgn2000}, it is constructed an exact solution of Einstein's field equations sourced by a scalar field assuming the flat rotation curve condition for the galactic halo. Taking into account this extended theory of gravity, it is worth notice that we have derived the same solution of \cite{mgn2000}, which is given by \eqref{sol2}. Furthermore, we introduced a geometric EoS of the form $w_{_{R\mathcal{G}}}  = - 1 - 2 / v_{tg}^{2}$. This gives us that the halo has always the property $w_{_{R\mathcal{G}}}  < -1$, which means that {\it the galactic halo should only be filled with phantom energy}, and therefore there is not exist the DM in the galactic halo for this extended theory of gravity.

----------------------------
\begin{figure*}[t]
\centering
   \includegraphics[scale=0.44]{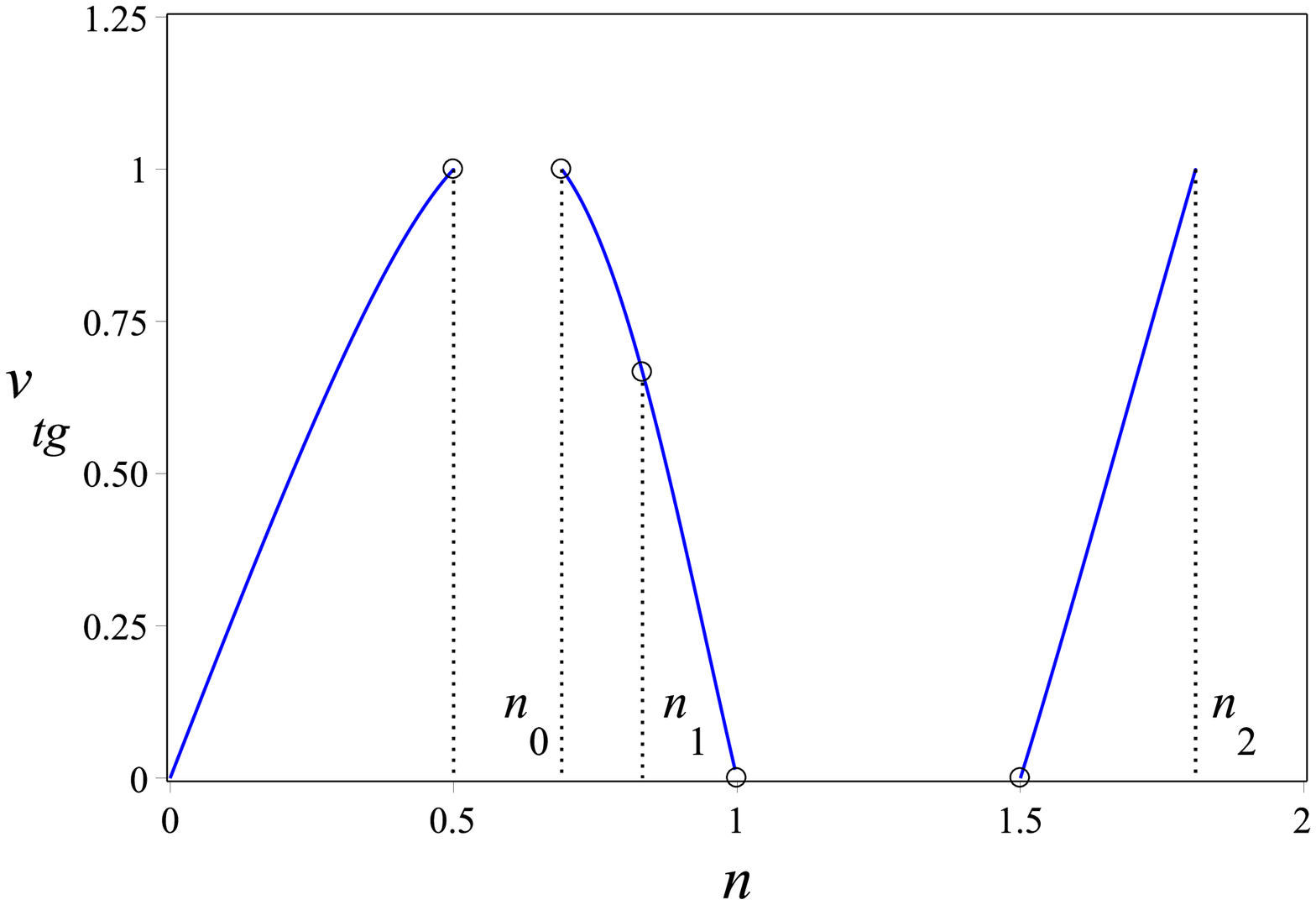}
   \includegraphics[scale=0.44]{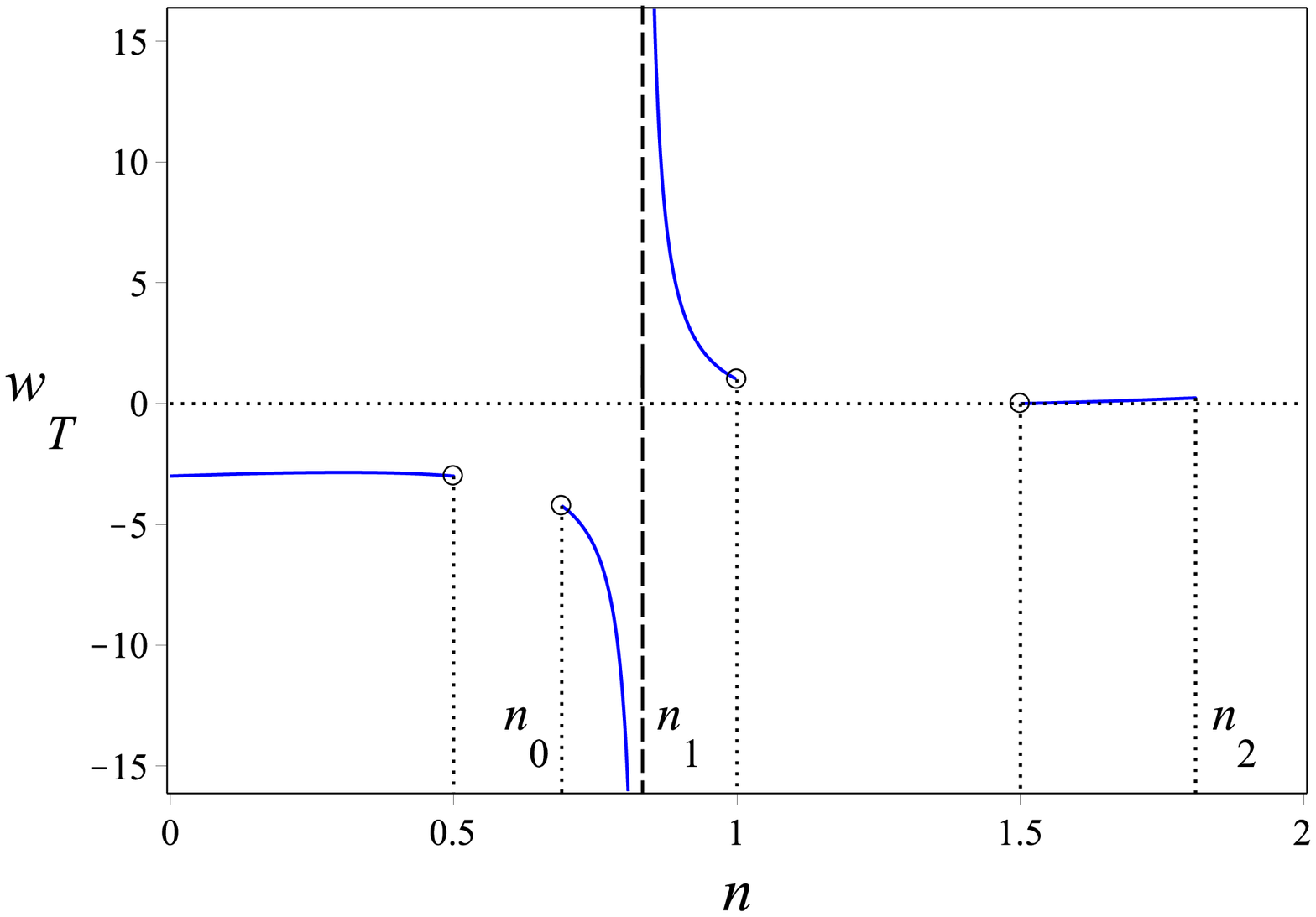}
   \includegraphics[scale=0.35]{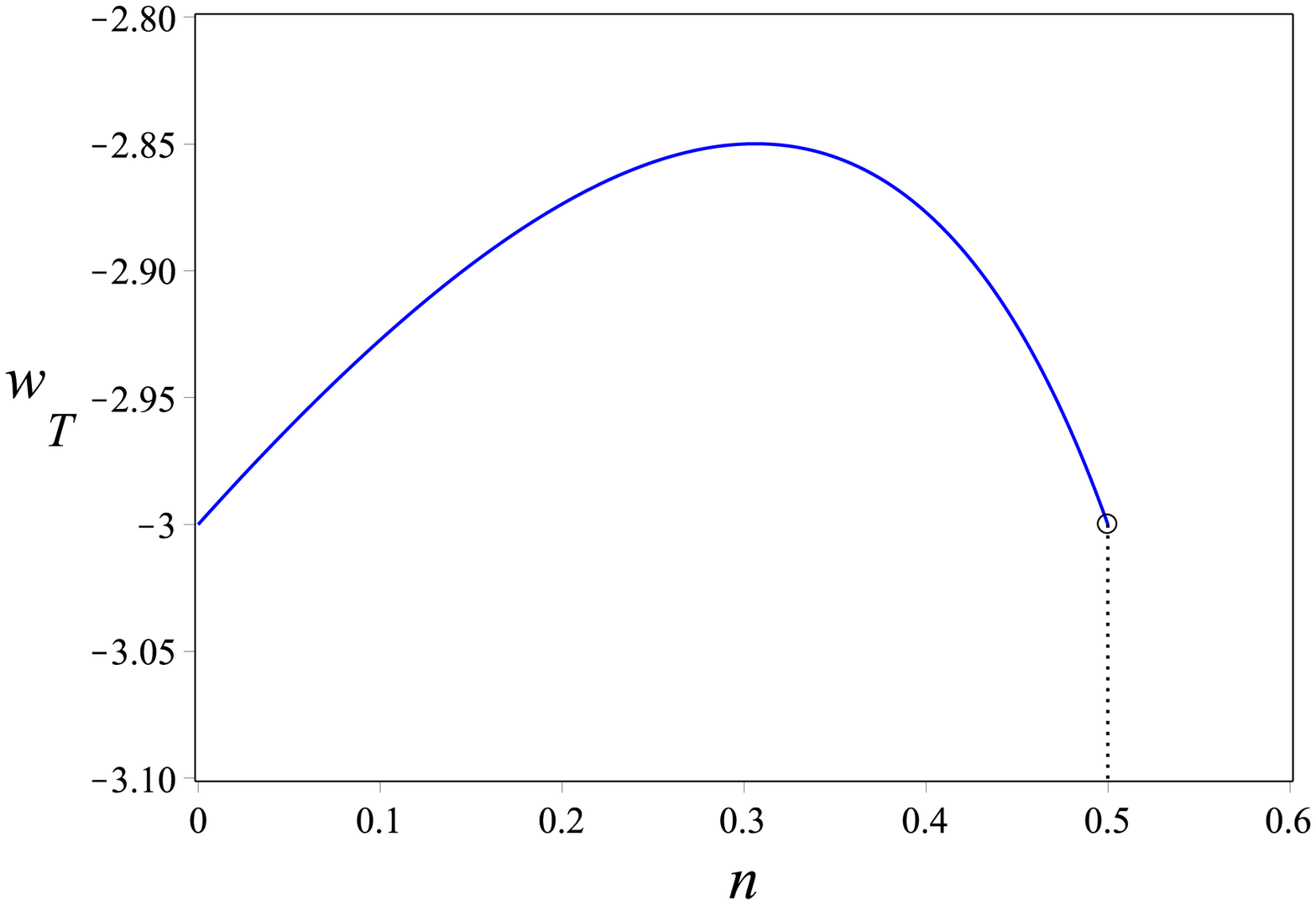}
   \includegraphics[scale=0.35]{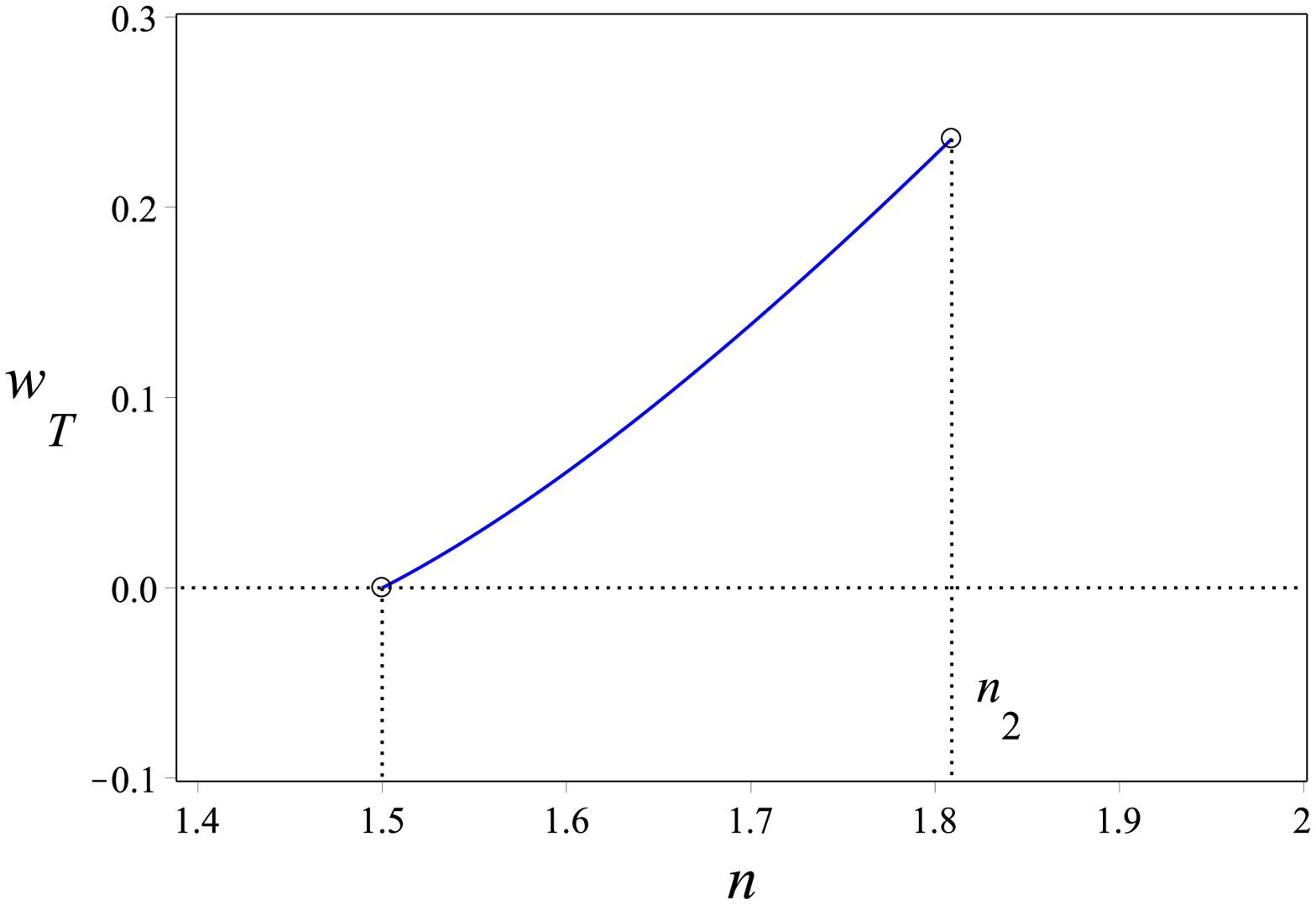}
\caption{ Plots showing the tangential velocity $v_{tg}$ (upper left panel), the Eq.\eqref{vtg-2}, and the geometrical equation of state $w_{_T}$ (upper right panel) given in \eqref{w-sol3} vs the power $n$ of $T^n$ gravity, where $n_0 = (5 - \sqrt{5})/4, \, n_1= 5/6$ and $n_2 = (5 + \sqrt{5})/4$. Since the shapes at the upper right panel are tiny in the ranges of $0 <n <0.5$ and $1.5 <n <n_2$, they are redrawn in the bottom two figures.
}
\label{fig2}
\end{figure*}
----------------------------

\item $f(T)= f_0 T^n$ gravity:

    For this theory of gravity, the relation between $v_{tg}$ and the power $n$, the Eq. \eqref{vtg-2}, is a cubic equation for $n$ as the form
\begin{equation}
n^3 - a_1 n^2 + a_2 n + a_3 = 0, \label{c-eq}
\end{equation}
    where $a_1 = \frac{1}{2} \left( v_{tg}^2 + 5 \right), \, a_2 = \frac{1}{2} \left( 2 v_{tg}^2 + 3 \right)$ and $a_3 = -\frac{5}{8} v_{tg}^2$. This cubic equation can be simplified by making the substitution $n = x + a_1 /3$. In terms of the new variable $x$, Eq. \eqref{c-eq} then becomes $x^3 + 3 P x - 2 Q = 0$, where $P = ( 3 a_2 - a_1^2)/9$ and $Q = ( 2 a_1^3 - 9 a_1 a_2 - 27 a_3) / 54$. Defining the polynomial discriminant $D = Q^2 + P^3$, we can solve algebraically the latter cubic equation. If $D > 0$, one of the roots is real and the other two roots are complex conjugates. If $D < 0$, all roots are real and unequal. In the latter case, defining $y = \arccos \left( Q / \sqrt{-P^3} \right)$, then the real valued solutions of \eqref{c-eq} are of the form $n_{k} = \frac{a_1}{3} + 2 \sqrt{-P} \cos \left( \frac{2 \pi k}{3} + \frac{y}{3} \right)$, where $k \in \{ 0,1,2\}$ and $P \leq 0$. In addition to the restrictions on $n$ such that $n \neq 0,1,\frac{1}{2},\frac{5}{6},\frac{5}{4},\frac{3}{2}$, we have additional property of $n$ due to $v_{tg} \in (0,1)$ as $n \in (0,\frac{1}{2}) \cup (n_0,n_1) \cup (n_1,1) \cup (\frac{3}{2},n_2)$, where $n_0, n_1$ and $n_2$ are given in the caption of Fig. \ref{fig2}. For the rotational velocity $v_{tg} = 10^{-3}$ of spiral galaxies, the Eq. \eqref{c-eq} has three real roots $n_1 = 4.169 \times 10^{-7}, \, n_2 = 0.99999975$ and $n_3= 1.500000333$, and the roots $n_2$ and $n_3$ have the property such that $w_{_T} = 1.000003$ for $n_2$, which is very close to unity, and $w_{_T} = 1.67\times 10^{-7}$ for $n_3$, which is very close to zero. If $n= 5/2$, then $w_{_T} = 1$ (stiff dark matter EoS parameter) and $v_{tg} = \sqrt{3}$ which means that it exceeds the speed of light. Further, one get the dark radiation EoS parameter, $w_{_T} = 1/3$, if $n = \frac{3}{2} \pm \frac{1}{\sqrt{6}}$ where both values are not in the interval of $n \in (0,\frac{1}{2}) \cup (n_0,n_1) \cup (n_1,1) \cup (\frac{3}{2},n_2)$. So we conclude that neither stiff DM nor dust DM or dark radiation exist at halos of galaxies in $T^n$ gravity. In Fig. \ref{fig2}, it is seen that $w_{_T} < -1$ for $n \in (0,\frac{1}{2}) \cup (n_0,n_1)$ which gives the phantom energy region of galactic halos, and  $w_{_T} >0$ for $n \in (n_1,1) \cup (\frac{3}{2},n_2)$.
    In Table 2, we present the tangential velocity of test particles, the corresponding power $n$, and the equation of state parameter $w_T$ for $T^n$ gravity by using the observed best-fit values of $v_{\rm flat}$ to the flat rotation curves of sample galaxies \cite{frank2016,cp1998,kamamda2017}.

    In Ref. \cite{fs2018}, a model with $f(T) = T + \alpha T^n$ gravity was used to reproduce galactic rotation curves, and found that in the $n \rightarrow 1$ limit the results were not approaching the case of GR. They have pointed out that the most promising region for $n$ is $1 < n < \frac{3}{2}$, but this range of $n$ is excluded in our work (see Fig.2). As claimed in Ref. \cite{bsz2020}, the work in Ref. \cite{fs2018} is problematic in the sense that they used an incorrect perturbed solution derived from an earlier paper \cite{rr2015}, where the authors used incorrect field equations in $f(T)$ gravity. In our study we have completely neglected the effect of the baryonic matter on the space-time geometry. So, it would be an interesting issue to analyze what could happen with including the baryonic matter to the field equations for $f(T)$ gravity.

\end{itemize}

\begin{table*}
\caption{\label{Tab2} Properties of sample galaxies in $T^n$ gravity. An explanation of the quantities appeared in the columns: $v_{\rm flat}$ - galactic flat rotation velocity data, $v_{\rm tg}$-tangential velocity, $c$-velocity of light, $n$-power of $T^n$, $w_T$- equation of state parameter. }
\begin{tabular}{ccccc}
Galaxy & $v_{\rm flat} ({\rm km \, s^{-1}})$  & $v_{\rm tg} = v_{\rm flat} / c$ ( $ 10^{-3} $ ) & $n$ & $w_T = p_1 / \rho_1$ \\
\hline
Milky Way & $200$ & $0.667$  & $0.999999888$ & $1.000001344$ \\ & & & $1.500000148$ & $7.4\times 10^{-8}$ \\  & & & $1.8537\times 10^{-7}$ & $-2.999999851$  \\ \\
NGC2841   & $310$ & $1.034$  & $0.9999997327$ & $1.000003208$ \\ & & & $1.500000356$ & $1.78\times 10^{-7}$ \\ & & & $4.4548\times 10^{-7}$ & $-2.999999644$  \\ \\
NGC3521   & $231$ & $0.77$   & $0.9999998614$ & $1.000001783$ \\ & & & $1.5000001783$ & $9.9\times 10^{-8}$ \\ & & & $2.4704\times 10^{-7}$ & $-2.999999803$  \\ \\
NGC3198   & $152$ & $0.507$  & $0.9999999357$ & $1.000000772$ \\ & & & $1.500000086$ & $4.3\times 10^{-8}$  \\ & & & $1.071\times 10^{-7}$ & $-2.999999916$  \\ \\
NGC2403   & $97$  & $0.324$  & $0.9999999738$ & $1.000000314$ \\ & & & $1.500000035$ & $1.75\times 10^{-8}$ \\ & & & $4.374\times 10^{-8}$ & $-2.999999965$  \\ \\
UGC128    & $140$ & $0.47$   & $0.9999999448$ & $1.000000662$ \\ & & & $1.500000074$ & $3.7\times 10^{-8}$  \\ & & & $9.204\times 10^{-8}$ & $-2.999999927$  \\ \\
UGC5721   & $80$  & $0.267$  & $0.9999999822$ & $1.000000214$ \\ & & & $1.500000024$ & $1.2\times 10^{-8}$  \\ & & & $2.9704\times 10^{-8}$ & $-2.999999977$  \\ \\
UGCA442   & $57$  & $0.19$   & $0.999999991$ & $1.000000108$ \\ & & & $1.500000012$ & $6\times 10^{-9}$      \\ & & & $1.5042\times 10^{-8}$ & $-2.999999988$  \\ \\
DDO154    & $45$  & $0.15$   & $0.9999999944$ & $1.000000067$  \\ & & & $1.500000007$ & $4\times 10^{-9}$  \\ & & & $9.375\times 10^{-9}$ & $-2.999999992$  \\
\hline
\end{tabular}
\end{table*}

\newpage

The cold dark matter (CDM) paradigm has been extremely successful in reproducing expansion history and large-scale structure of the universe as well as the observed DM halos of galaxies. In a CDM model, the DM in the universe is arranged in DM halos of galaxies. It would be mentioned a possibility that the CDM paradigm may break down on galactic scales. Supposing the DM particles are warm, instead of cold, then it gives rise that they were quasi-relativistic during kinetic decoupling from the thermal bath in the early universe \cite{bot2001}. The results of this work are valid for either cold or warm DMs.

In our approach, we need to mention that the energy density $\rho_X$ and the pressure $p_X$, where $X \in \{ R, \mathcal{G}, T\}$, are effective geometric quantities. So, these quantities may not have an exact physical interpretation as for the similar physical fields.
In this work, the modified gravity theories such as $R^n, T^n$ and $f(R,\mathcal{G}) = f_1 R + f_2 \sqrt{\mathcal{G}}$ have been introduced as a possible way to explain the observed flat rotational velocities of galaxies without the need of any DM component. It is seen from the above three examples that the $R^n$ and $T^n$ gravities have more rich structures than the $f(R,\mathcal{G})=f_1 R + f_2 \sqrt{\mathcal{G}}$ gravity. For future aim at improving the halo model discussed here, it would be complementary to take into account the ordinary matter in addition to the DM contribution in the halos of galaxies.

\newpage



\end{document}